\documentclass[amsmath,prb,twocolumn,superscriptaddress]{revtex4-1}
\usepackage{color}
\usepackage{amsfonts}
\usepackage{amssymb}
\usepackage{graphicx}
\usepackage{pstricks}

\begin{document}

\title{Local-noise spectroscopy for non-equilibrium systems}
\author{Gabriel Cabra}
\email{gcabra@ucsd.edu}
\affiliation{Department of Chemistry \& Biochemistry, University of California San Diego, La Jolla, CA 92093, USA}
\author{Massimiliano Di Ventra}
\email{diventra@physics.ucsd.edu}
\affiliation{Department of Physics, University of California San Diego, La Jolla, CA 92093, USA}
\author{Michael Galperin}
\email{migalperin@ucsd.edu}
\affiliation{Department of Chemistry \& Biochemistry, University of California San Diego, La Jolla, CA 92093, USA}



\begin{abstract}
We introduce the notion, and develop the theory of local-noise spectroscopy (LNS) - a tool to study the 
properties of systems far from equilibrium by means of flux 
density 
correlations. As a test bed, we apply it to biased molecular junctions. This tool naturally extends those based on local fluxes,
while providing complementary information on the system.
As examples of the rich phenomenology that one can study with this approach, we show that LNS can be used to yield information on microscopic 
properties of bias-induced light emission in junctions, provide local resolution of intra-system interactions, and employed as a nano-thermometry tool. Although LNS may, at the moment, be difficult to realize experimentally, it can nonetheless be used 
as a powerful theoretical tool to infer a wide range of physical properties on a variety of systems of present interest.
\end{abstract}
\maketitle
\section{Introduction}\label{intro}
Local spectroscopic tools, such as scanning tunneling microscopy (STM)~\cite{binnig1982surface} or atomic force microscopy (AFM)~\cite{binnig1986atomic}, have long been used to study the physical properties 
of a wide variety of physical systems. Recent developments have added further capabilities and pushed the resolution of spectroscopic techniques even further. For instance, STM has been employed in imaging with sub-molecular spatial 
resolution~\cite{BredeWiesendangerMRS14,HoPRL15,HoJPCC15,DongJapJApplPhys15} (including specific vibrational mode imaging~\cite{HoJCP11}, bond-selective chemistry~\cite{HoNatChem12}, and spatial distribution of additional charges on molecules~\cite{ReppNL11}), while AFM was utilized to study single-electron transfer between molecules~\cite{MeyerNatComm15}, and to resolve intra-molecular structures~\cite{JarcvisIntJMolSci15}. 

Additional spectroscopic tools that have also been employed in non-equilibrium conditions include surface- and 
tip-enhanced Raman spectroscopy which now allows for measurements on the Angstr\"om scale~\cite{ApkarianACSNano17,JensenApkarianACSNano17,JensenACSNano17,ElKhouryNL15,ElKhouryChemComm17}, studies of 
electronic pathways in redox proteins~\cite{GorostizaSmall17}, utilizing 
3D dynamic probe of single-molecule 
conductance to explore conformational changes~\cite{ShigekawaNatComm15}, and energy-resolved atomic probes~\cite{Erasp}, to name just a few. 

The majority of these techniques employ inter-atomic 
fluxes (currents) to extract information on the physical system. On the other hand, fluctuations of the current (noise) typically provide complementary information to flux measurements~\cite{Di_Ventra_book}. 
For example, shot noise at biased nanoscale junctions yields information on the number of scattering 
channels~\cite{RuitenbeekNL06},
effects of intra-molecular interactions on transport~\cite{YeyatiRutenbeekPRL12}, 
and effective charge of carriers~\cite{RonenPNAS16}. 
Connection between current-induced light emission and electronic 
shot noise was demonstrated experimentally~\cite{BerndtPRL12}, 
and a theory of light emission from quantum noise was formulated~\cite{KaasbjergNitzanPRL15}. 
Recently, shot noise was employed to image 
hot electron energy dissipation~\cite{ZhenghuaWeiScience18}, and extract information on the local electronic 
temperatures~\cite{Natnoise14,KhrapaiSciRep16}.

All these experiments 
typically deal with mesoscopic (or macroscopic) regions of the samples, hence do not really provide direct information on the local properties of the system. Here instead we introduce and develop the theory of {\it local noise spectroscopy} (LNS) for non-equilibrium systems, and show that it can yield information on a wide range of physical properties otherwise difficult to obtain with other means. 

To illustrate the proposed LNS approach we apply it to biased molecular junctions that have been widely studied in a variety of contexts~\cite{NitzanRatnerScience03,visionsNatNano13}. 
We then show that LNS can be used to extract microscopic properties of bias-induced light emission in 
molecular junctions, 
and provides local resolution of intra-system interactions revealing the relevant energy scale(s)  
in coherent quantum transport.  
We further discuss the LNS application to yet another property: nano-thermometry. 

Of course, ``locality'' is strongly related to the size of the surface area of the experimental probe(s) that need to be coupled to the system to extract the necessary quantities. At the moment, probes with the resolution that we discuss in this paper are difficult to realize. Nonetheless, we hope that by showing the rich physical information that LNS can provide on a wide range of systems may motivate experimental studies in this direction.


The structure of the paper is as follows. Section~\ref{model} introduces the model of a junction and yields theoretical 
details of local noise simulations. Results of the simulations and discussion are described in 
Section~\ref{numres}. We summarize our findings and indicate directions for future research
in Section~\ref{conclude}.


\section{\label{model}Local Noise Spectroscopy}
We follow the work reported in Ref.~\cite{CabraJensenMGJCP18} and consider 
a nanoscale system (a molecule) $M$ coupled to two macroscopic contacts $L$ and $R$, 
each at its own local equilibrium. Difference in electrochemical potentials on the two contacts causes electron flux through the molecule. The Hamiltonian of the system is 
\begin{equation}
\label{H}
 \hat H = \hat H_M + \sum_{K=L,R}\big(\hat H_K+\hat V_{KM}\big),
\end{equation}   
and consists of molecular, 
\begin{align}\label{HM}
\hat H_M &= \sum_{m_1,m_2\in M} H_{m_1m_2}^{M}\hat d_{m_1}^\dagger\hat d_{m_2}
\nonumber \\ &+ \frac{1}{2}\sum_{\begin{subarray}{c}m_1,m_2\\m_3,m_4\end{subarray}\in M} V_{m_1m_2,m_3m_4}^{M}
\hat d_{m_1}^\dagger\hat d_{m_2}^\dagger\hat d_{m_4}\hat d_{m_3},
\end{align}
and contacts, $\hat H_{K} = \sum_{k\in K} \varepsilon_k\hat c_k^\dagger\hat c_k$, components. 
$\hat V_{KM}=\sum_{m\in M}\sum_{k\in K}\big(V_{mk} \hat d_m^\dagger\hat c_k + H.c.\big)$
describes the electron transfer between the molecule and the contacts.  
Here, $\hat d_m^\dagger$ ($\hat d_m$) and $\hat c_k^\dagger$ ($\hat c_k$)
creates (destroys) an electron in single-particle states $m$ of the molecule and $k$ of the contacts,
respectively. The first term on the right-hand side of Eq.~(\ref{HM}) represents the kinetic energy of electrons
and part of their potential energy due to interaction with static nuclei and external fields; the 
second term introduces electron-electron interactions (assumed to be confined to the molecular subspace). 

The current-density operator is~\cite{Landau_v3_1991} 
(here and below $e=\hbar=m=1$)
\begin{equation}
\label{cur_op}
 \hat{\vec j}(\vec r) = -\frac{i}{2}\bigg(
 \vec\nabla \hat\psi^\dagger(\vec r)\, \hat\psi(\vec r) 
 - \hat\psi^\dagger(\vec r)\,\vec\nabla \hat\psi(\vec r)
 \bigg),
\end{equation}
where $\hat\psi^\dagger(\vec r)$ ($\hat\psi(\vec r)$) is the field operator creating (annihilating) an 
electron at position $\vec r$. Within the molecular subspace, spanned by a basis $\{\phi_m(\vec r)\}$, the field operator can be expanded as $\hat\psi(\vec r)=\sum_{m} \hat d_m\phi_m(\vec r)$. 
The current density is then~\cite{Di_Ventra_book}
\begin{align}
\label{current}
 \vec j(\vec r,t) &\equiv \mbox{Tr}\big[\hat{\vec j}(\vec r)\,\hat \rho(t)\big]
\nonumber \\
 &= -\frac{1}{2}\sum_{m_1,m_2\in M} G^{<}_{m_1m_2}(t,t)\,d\vec A_{m_2m_1}(\vec r),
\end{align}
where $\hat\rho(t)$ is the system's density operator,  
\begin{equation}
d\vec A_{m_2m_1}\equiv \vec\nabla\phi_{m_2}^{*}(\vec r)\,\phi_{m_1}(\vec r) - \phi_{m_2}^{*}(\vec r)\,\vec\nabla\phi_{m_1}(\vec r),
\end{equation}
and $G^{<}_{m_1m_2}(t,t)$ is the equal-time lesser projection
of the single-particle Green's function
\begin{equation}
\label{Gdef}
 G_{m_1m_2}(\tau_1,\tau_2) \equiv 
 -i\big\langle T_c\,\hat d_m(\tau_1)\,\hat d_{m_2}^\dagger(\tau_2)\big\rangle.
\end{equation}
The symbol $T_c$ represents the Keldysh contour ordering operator, $\tau_{1,2}$ are the contour variables, and the average is taken over the initial-time density 
operator.
Eq.~(\ref{current}) was used in previous studies of current density in nanoscale 
junctions~\cite{XueRatnerPRB04,EversPRL14,EversPRB15,EversJCTC15,NozakiSchmidtJCC17,CabraJensenMGJCP18}.

In order to compute the local noise properties of the system, we consider the local current-current correlation function~\cite{kubo1957statistical}
\begin{equation}
\label{corr}
C_{i_1i_2}^{j}(\vec r_1,t_1;\vec r_2,t_2) \equiv 
\big\langle \delta\hat{j}_{i_1}(\vec r_1,t_1)\,\delta\hat{j}_{i_2}(\vec r_2,t_2) \big\rangle .
\end{equation}
Here, operators are in the Heisenberg picture, $i_{1,2}\in\{x,y,z\}$,
$\langle\ldots\rangle\equiv\mbox{Tr}[\ldots \hat\rho_0]$,
and $\delta\hat{j}_i(\vec r) \equiv \hat{j}_i(\vec r) - \langle \hat{j}_i(\vec r) \rangle$.
In terms of the correlation functions, the local noise is then 
\begin{align}
\label{loc_noise}
 &S_{i_1i_2}(\vec r_1,\vec r_2;t_1,t_2) =
 \\
 & \frac{1}{4}\bigg( 
    C_{i_1i_2}^{j}(\vec r_1,t_1;\vec r_2,t_2) + C_{i_2i_1}^{j}(\vec r_2,t_1;\vec r_1,t_2)
  \nonumber   \\ &
 + C_{i_1i_2}^{j}(\vec r_1,t_2;\vec r_2,t_1) + C_{i_2i_1}^{j}(\vec r_2,t_2;\vec r_1,t_1) 
 \bigg).
 \nonumber
\end{align}

Of course, in realistic settings, the current density needs to be averaged over a surface area $A$, which determines the actual resolution of this local noise spectroscopic probe. Since two current densities appear in Eq.~(\ref{corr}), we need to choose two surface areas with corresponding orientations 
\begin{equation}\label{loc_noise_A}
S_{A_1A_2}(t_1,t_2)
=\int_{A_1}d\vec s_1\int_{A_2} d\vec s_2\, S_{i_1i_2}(\vec r_1,\vec r_2;t_1,t_2),
\end{equation}
where $d\vec s_1$ and $d\vec s_2$ are two infinitesimal surfaces whose normal orientation is parallel to the directions $i_1$ and $i_2$, respectively. 
Note that the local noise, Eq~(\ref{loc_noise}), and the integrated noise, Eq.~(\ref{loc_noise_A}),
matrices are Hermitian: $\mathbf{S}\equiv \mathbf{S}^\dagger$. 

Equation~(\ref{loc_noise_A}) 
allows the computation of several properties, both at steady state and not. In addition, it is general: it is valid when the system is both close to equilibrium and far from it, in the presence of weak or strong interactions. 

As illustration, below we focus only on steady-state properties, and treat the electron-electron interaction at the mean-field (Hartree-Fock) level.  
We note that although we consider a non-interacting (mean-field) model and focus on the steady-state situation,
the theory can be extended to time-dependent and interacting systems.
For the description of transient processes (such as those considered, e.g. in Ref.~\onlinecite{feng_current_2008}) in noninteracting systems 
one has to simulate time-dependent single-particle Green's functions as done, e.g., in Ref.~\onlinecite{sukharev_transport_2010}. 
Weak interactions, where perturbation theory can be applied, can be treated in a similar 
manner as in Refs.~\onlinecite{souza_spin-polarized_2008,myohanen_kadanoff-baym_2009}.
In the case of strong interactions the situation becomes much more complicated, and numerically-heavy methods are required
in this case~\cite{ridley_numerically_2018}. At present, such methods are restricted to simple models only.

Note also that standard zero-frequency shot noise (as well as the noise spectrum)~\cite{BlanterButtikerPR00} 
can be obtained straightforwardly from our expressions by performing integration in Eq.~(\ref{loc_noise_A}) 
over surfaces separating the molecule from the contacts. 
Indeed, the integral of each local flux over such surfaces by definition yields the total current flowing between the molecule and the contact. 
Therefore, one obtains the current-current (more precisely current fluctuation-current fluctuation) correlation function, 
which is the standard definition of noise at the molecule-contact interface.

At steady state, it is convenient to consider the Fourier transform of the expression (\ref{loc_noise}). In addition, the single-particle (mean-field) level of description allows us to simplify the LNS expression via the use of Wick's theorem~\cite{Mahan_1990}. 
Under these conditions the local noise expression (\ref{loc_noise}) becomes
\begin{align}
\label{loc_noise_w}
& S_{i_1i_2}(\vec r_1,\vec r_2;\omega) \equiv 
 2 \int_{-\infty}^{+\infty}dt\, e^{i\omega t} S_{i_1i_2}(\vec r_1,\vec r_2;t)
 \\
 &=  \frac{1}{8}\sum_{\begin{subarray}{c}m_1,m_2\\m_3,m_4\end{subarray}\in M}
 d[A_{m_2m_1}]_{i_1}\, d[A_{m_4m_3}]_{i_2}
 \int_{-\infty}^{+\infty}\frac{dE}{2\pi}
 \nonumber \\ &
 \bigg(\;
    G^{>}_{m_1m_4}(E+\omega)\, G^{<}_{m_3m_2}(E) 
 + G^{>}_{m_3m_2}(E+\omega)\, G^{<}_{m_1m_4}(E)
\nonumber \\ &
 + G^{>}_{m_1m_4}(E-\omega)\, G^{<}_{m_3m_2}(E)
 + G^{>}_{m_3m_2}(E-\omega)\, G^{<}_{m_1m_4}(E)
 \bigg),
 \nonumber
\end{align}
where $G^{<(>)}_{mm'}(E)$ is the Fourier transform of lesser (greater) projection of
the single-particle Green's function (\ref{Gdef}).
\begin{figure}[t]
	\centering\includegraphics[width=\linewidth]{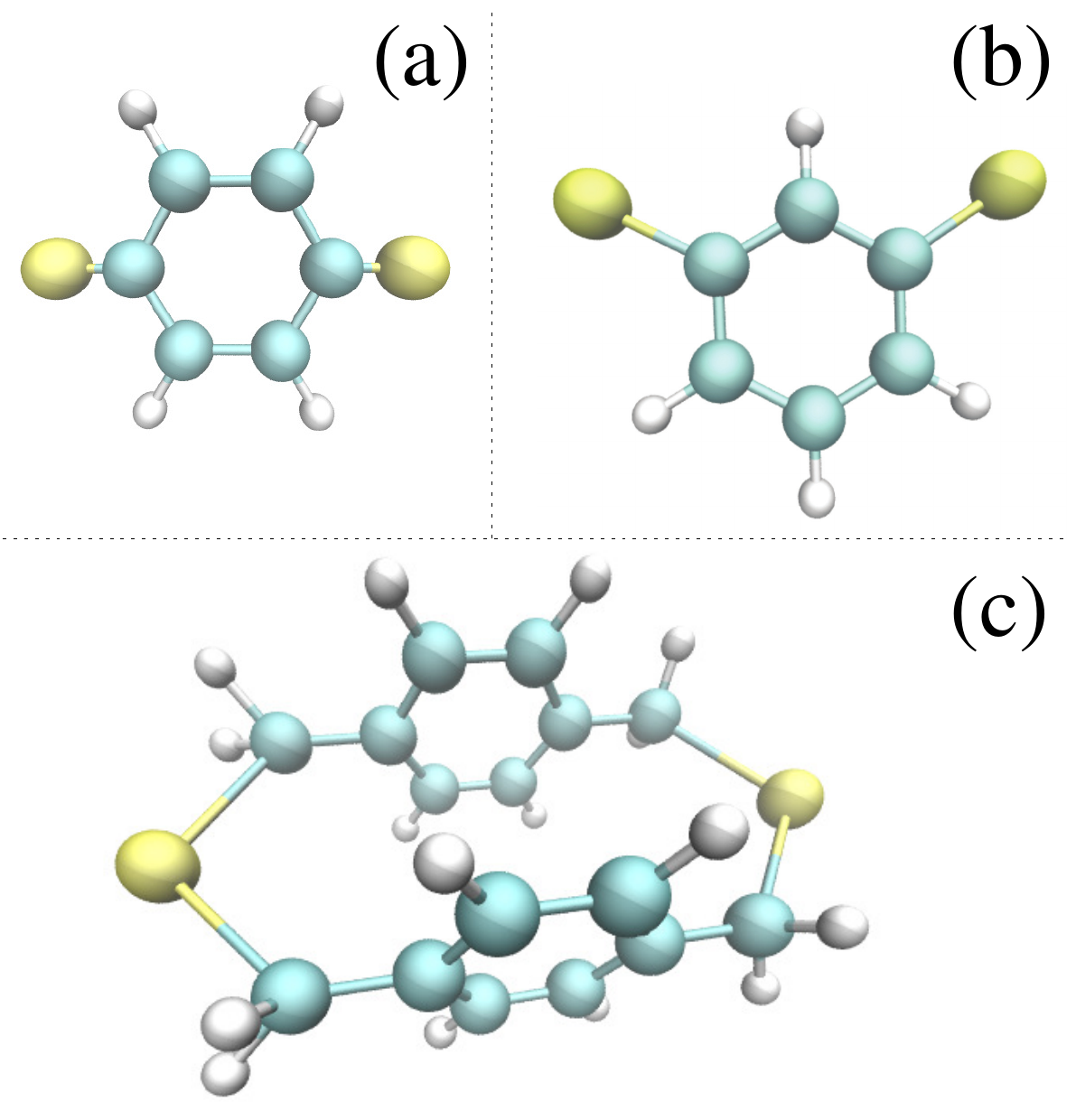}
	\caption{\label{fig1}
		Molecular junctions used to illustrate the local-noise spectroscopy: 
		(a) para-benzenedithiol (PBDT),
		(a) meta-benzenedithiol (MBDT), and
		(b) 2,11-dithi(3,3)paracyclophane molecular structures. 
		The yellow circles represent the sulfur atoms that attach to two macroscopic electrodes.
	}
\end{figure}

\section{\label{numres}Numerical results}
We are now ready to illustrate how the LNS, implemented in Eq.~(\ref{loc_noise_w}), may be used to analyze several physical properties 
in biased molecular junctions. 
We then consider three distinct molecular structures represented in Fig.~\ref{fig1}: a benzenedithiol molecular junction in 
para (PBDT, Fig.~\ref{fig1}a) and meta (MBDT, Fig.~\ref{fig1}b) configuration, and a 2,11-dithia(3,3)paracyclophane molecular
junction employed in measurements of quantum coherence in Ref.~\cite{VenkataramanHybertsenNatNano12} (Fig.~\ref{fig1}c). Simulations of molecular electronic
structure are performed within the Gaussian package~\cite{g09} employing the Hartree-Fock level of the 
theory, and Slater-type orbitals with 3 primitive gaussians (STO-3g) basis set. 
The molecular structures are coupled to semi-infinite contacts via sulfur atoms;
each orbital of the latter is assumed to support $\Gamma_{K}=0.1$~eV electron exchange
rate between the molecule and contact $K$ ($L$ or $R$)~\cite{kinoshita_electronic_1995}. 
The contacts are modeled within the wide-band approximation~\cite{CabraJensenMGJCP18}.
While this level of molecule-contacts modeling is enough for illustration purposes, 
actual {\it ab initio} simulations should use a better basis set and perform realistic self-energy calculations.
The Fermi energy $E_F$ is taken to be $1$~eV above the highest occupied molecular orbital (HOMO)
for the PBDT and MBDT junctions (Figs.~\ref{fig1}a and b). 

Following Ref.~\cite{VenkataramanHybertsenNatNano12}
we take the Fermi energy in the middle of the highest occupied-lowest unoccupied molecular orbital
(HOMO-LUMO) gap for the double-backbone molecular structure of 
Fig.~\ref{fig1}c. 
Finally, the bias $V_{sd}$ across the junction  is applied symmetrically: $\mu_{L,R}=E_F\pm |e|V_{sd}/2$. 
The numerical illustrations below are presented on plane(s) parallel to the molecular plane(s)
at a distance of $1.5$~\AA\/ above it. Calculations are performed on a spatial grid
spanning from  $-4$~\AA\/ to $4$~\AA\/ with step of $0.25$~\AA. The center of the coordinate system is chosen at the molecule's center of mass.

\begin{figure}[t]
	\centering\includegraphics[width=6cm]{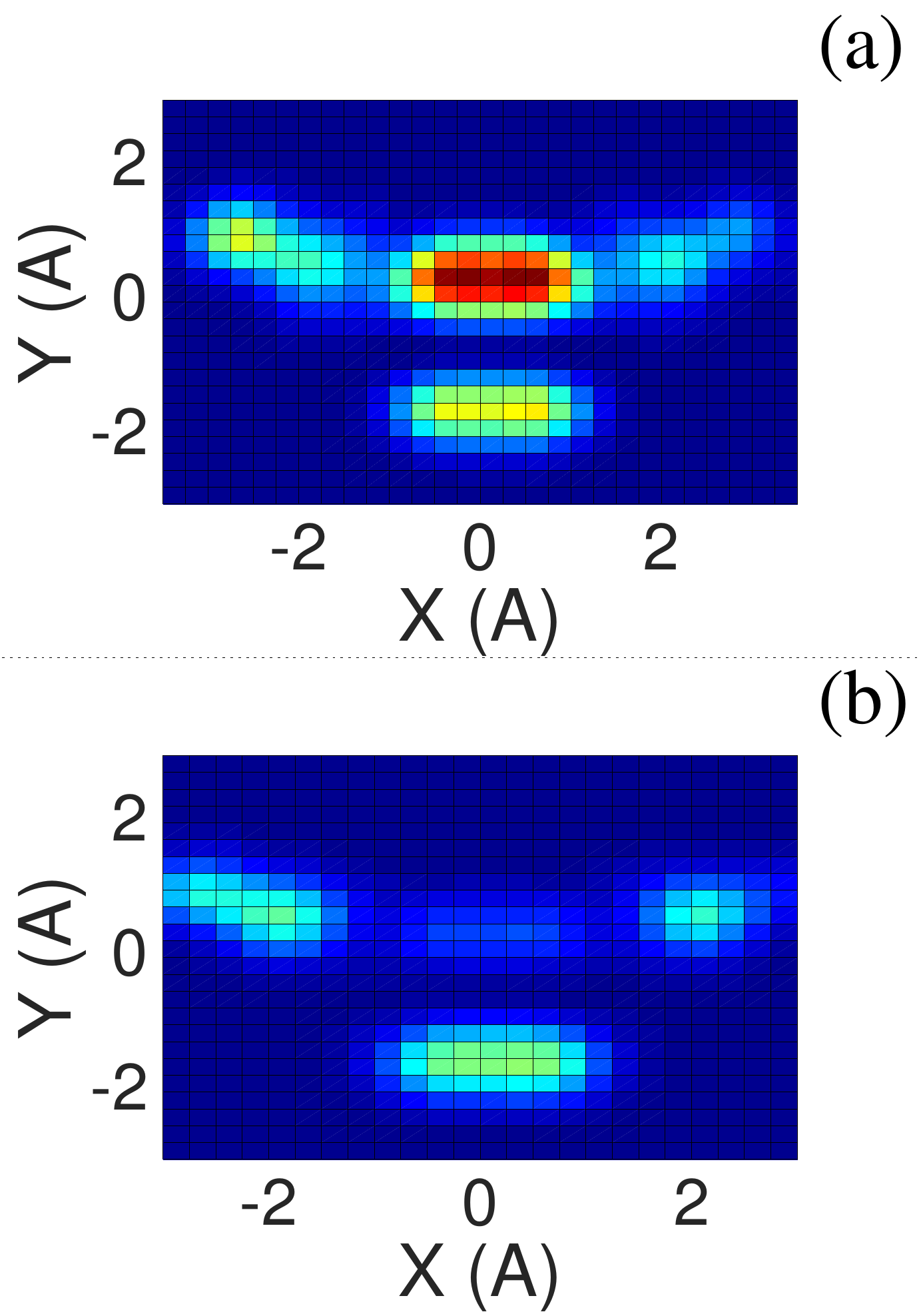}
	\caption{\label{fig2}
		Electroluminiscence profile in a MBDT molecular junction (Fig.~\ref{fig1}b) at $V_{sd}=3$~V. 
		The profile is plotted parallel to the molecule at $1.5$~\AA\/ above the molecular plane. 
		The horizontal axis ($X$) is the direction of tunneling in the junction. 
		The localized surface plasmon-polariton vector is assumed to be directed along $X$. 
		The light emission profiles are shown for the outgoing photons of frequencies
		(a) $\omega=2$~eV and (b) $3$~eV.
	}
\end{figure}

\begin{figure}[t]
	\centering\includegraphics[width=6cm]{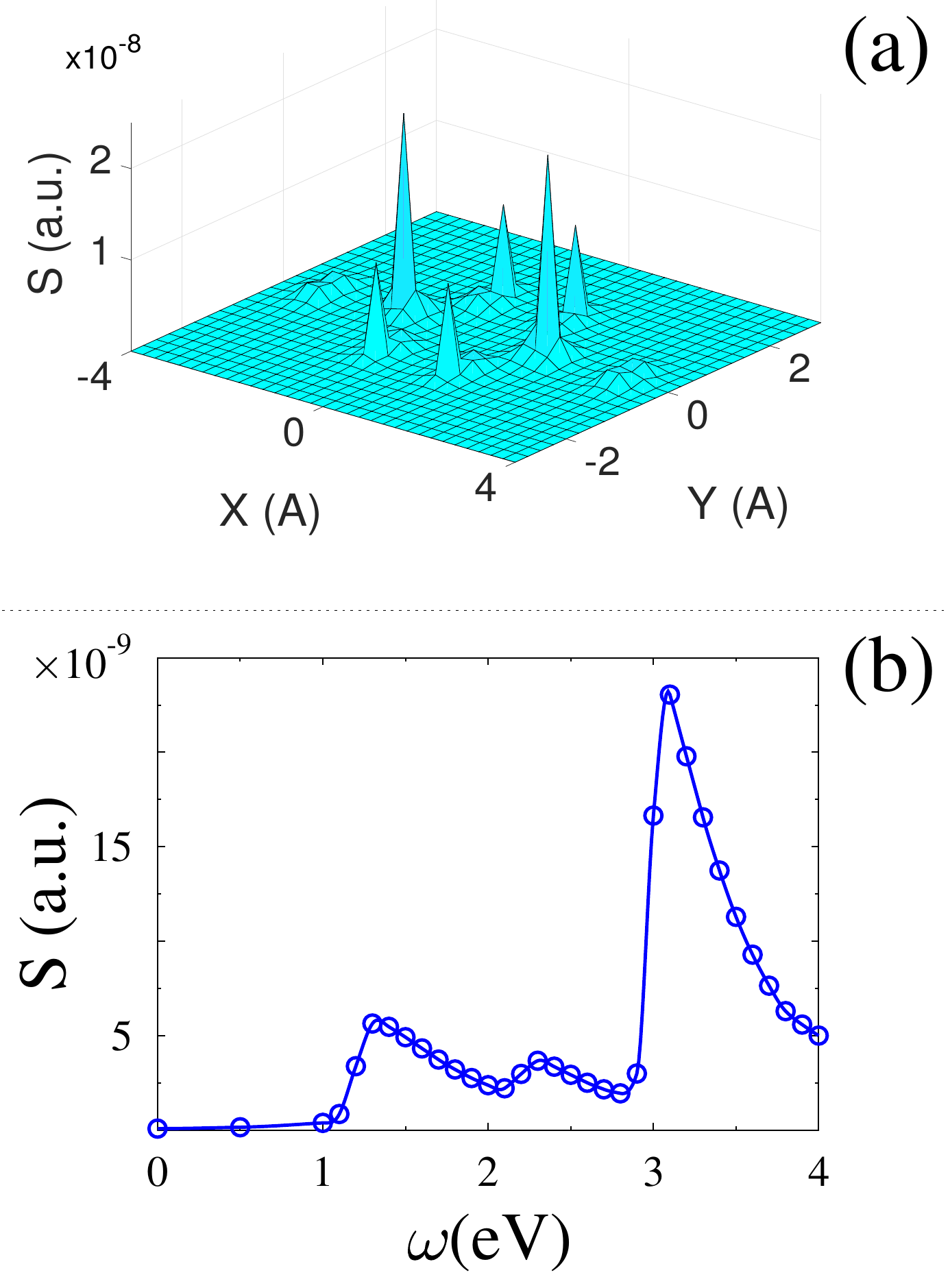}
	\caption{\label{fig3}
		Local noise cross-correlation 
		$S(\vec r_1,\vec r_2;\omega)=\sqrt{\sum_{i=\{x,y,z\}} S^2_{ii}(\vec r_1,\vec r_2;\omega)}$ 
		in a 2,11-dithi(3,3)paracylophane molecular junction (Fig.~\ref{fig1}c) at $V_{sd}=1$~V. 
		The horizontal axis ($X$) is the direction of tunneling in the junction. 
		The profile is plotted parallel to the molecular planes with $Z$ projections of $\vec r_1$
		and $\vec r_2$ taken $1.5$~\AA\/ on outer sides of molecular planes. 
		We show 
		(a) the cross-correlation map in the $XY$ plane at $\omega=3.5$~eV, and 
		(b) the cross-correlation at the points corresponding to positions of carbon atoms of the benzene rings
		vs. frequency $\omega$.
	}
\end{figure}

\begin{figure*}[htbp]
	\centering\includegraphics[width=14cm]{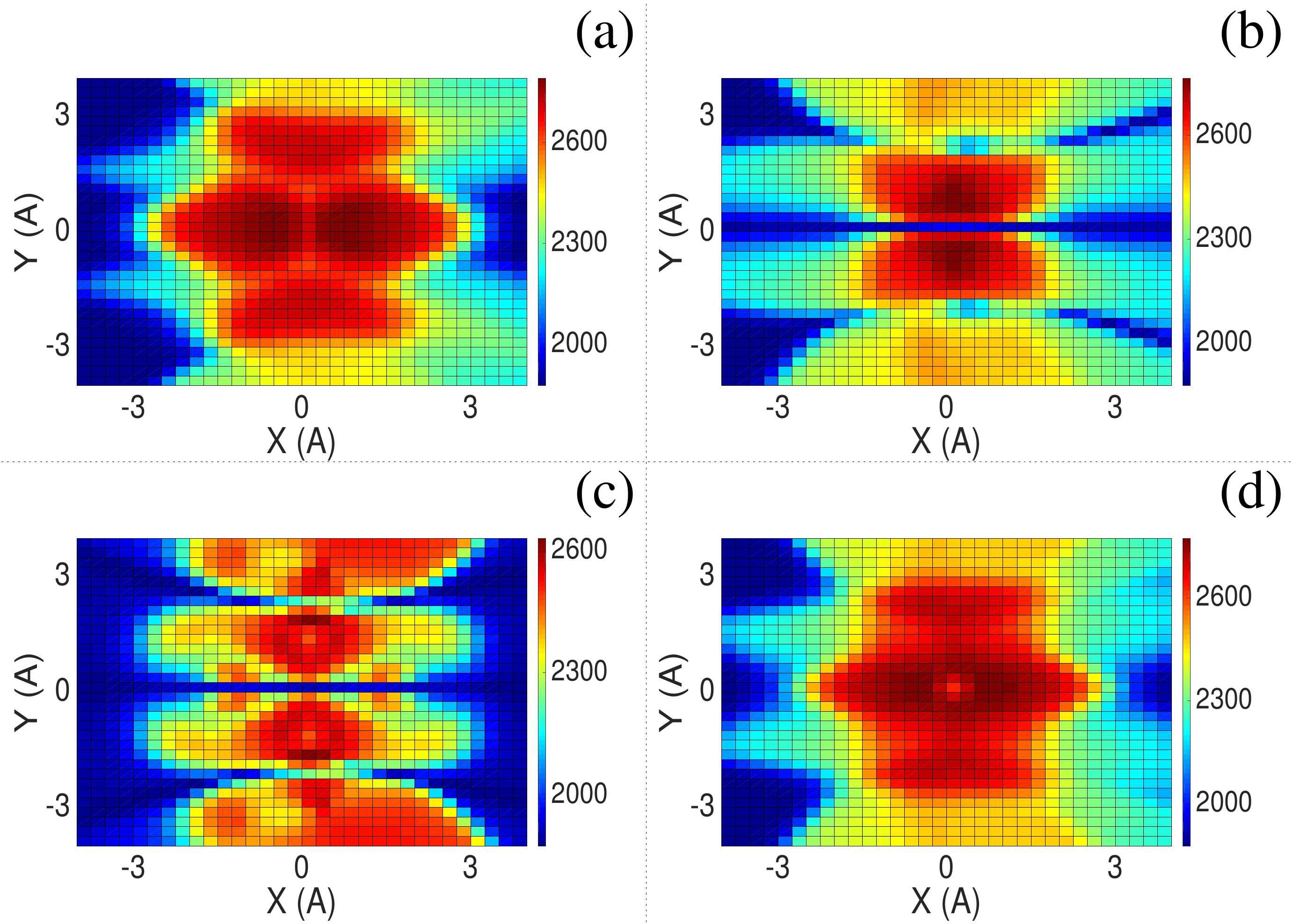}
	\caption{\label{fig2_SM}
		Local noise thermometry in a PBDT molecular junction (Fig.~\ref{fig1}a) at a bias $V_{sd}=1$~V. 
		The temperature distribution, Eq.~(\ref{TM}), is plotted parallel to the molecule at $1.5$~\AA\/ 
		above the molecular plane. The horizontal axis ($X$) is the direction of tunneling in the junction. 
		The temperature estimates are shown from a non-invasive probe oriented 
		(a) perpendicular to the junction tunneling direction,
		(b) parallel to the junction tunneling direction and perpendicular to the molecular plane,
		(c) parallel to the molecular plane, and
		(d) perpendicular to the local current. The vertical (color) bar shows the temperature scale in K. 
	}
\end{figure*}

{\it Bias-induced light-emission --} Let us first discuss bias-induced light emission in molecular junctions. 
The theory of light emission from quantum noise was recently put forward in Ref.~\cite{KaasbjergNitzanPRL15}.
It was shown that the emission is related to the positive frequency part of the asymmetric noise (the 
last row of Eq.~(\ref{loc_noise_w})) in the plasmonic contact. The corresponding local noise distribution then yields the electroluminscence profile in a biased molecular junction. According to the theory of Ref.~\cite{KaasbjergNitzanPRL15}, the 
local current projections are 
fixed by the direction of the localized  surface plasmon-polariton vector. 

Figure~\ref{fig2} shows the light emission in a MBDT molecular junction (Fig.~\ref{fig1}b)
at a bias $V_{sd}=3$~V. It is interesting to note that the outgoing photons of different frequencies
probe different parts of the molecule (compare Figs.~\ref{fig2}a and b). This fact cannot be extracted from other spectroscopic probes and is due to local features of the potential profile distribution in the junction.

{\it Correlation effects in transport --} As a second example, we discuss the local noise probed at two distinct points in space (cross-correlations) to detect 
inter-dependence of different paths in quantum transport. Using the latter as, e.g., indicator of inter-species
spin interactions was discussed in Ref.~\cite{Pershin16}. 
To illustrate the usefulness of the concept in a non-equilibrium setting we consider a 
2,11-dithi(3,3)paracylophane molecular junction (Fig.~\ref{fig1}c).
The junction provides two paths for electron tunneling, which lead to observation of constructive
interference in transport~\cite{VenkataramanHybertsenNatNano12}.

We probe the independence of the two paths by calculating the LNS cross-correlation map of local currents 
taken outside of the two molecules at a distance of $1.5$~\AA\/ away from the molecular planes.
That is, $X$ and $Y$ projections ($XY$ is parallel to the molecular planes) of vectors 
$\vec r_1$ and $\vec r_2$ in (\ref{loc_noise_w}) are taken equal to each other,
while their $Z$ components are taken $1.5$~\AA\/ away on the outer side of molecular planes
(see Fig.~\ref{fig1}c). The resulting cross-correlation map is shown in Fig.~\ref{fig3}.
As expected, correlations between the two molecules show maxima at the positions of carbon atoms,
where interaction between $p_z$ atomic orbitals of the atoms is significant
(see Fig.~\ref{fig3}a). 

However, extra information can be extracted from the frequency dependence of the cross-correlation, which indicates a characteristic  interaction energy scale in the system. 
Frequency dependence of cross-correlation corresponding to position of carbon atoms of the two 
benzene rings is shown in Fig.~\ref{fig3}b. Three peaks indicate respectively 
the effective strengths of $2s-2s$, $2s-2p_z$, and $2p_z-2p_z$ atomic orbital couplings.
The peaks approximately correspond to the Rabi frequency related to the Fock matrix couplings 
between orbitals of adjacent carbon atoms in the two molecules of the junction.

{\it Local thermometry --}
We discuss here how to employ local-noise spectroscopy (LNS) as a local thermometry 
tool in current-carrying molecular junctions. 
Although a nonequilibrium state cannot be identified with a unique thermodynamic 
temperature, and experimentally measurable failures of attempts to introduce such characteristic
were discussed in the literature~\cite{HartmannEPL05}, the 
concept of temperature as a single parameter effectively describing bias-induced heating~\cite{DiVentraTaoNL06} 
is attractive. For example, Raman measurements in current-carrying junctions were utilized
to introduce an effective temperature of molecular vibrational and electronic degrees
freedom~\cite{CheshnovskySelzerNatNano08,NatelsonNatNano11}.
Such assignment implies existence of some sort of local equilibrium. As a measure of electronic temperature
an idea of equilibrium probe with chemical potential and temperature adjusted in such a way
that no particle and energy fluxes exist between the probe and nonequilibrium electronic distribution
was put forward and utilized in a number of studies~\cite{DiVentraRMP11,MGANJPCL11,MGANPRB11}.

In the context of noise spectroscopy, it was indeed recently suggested that noise may be used to measure electronic
temperatures~\cite{KhrapaiSciRep16,ZhenghuaWeiScience18}. Here, we follow the 
suggestion of Ref.~\cite{ZhenghuaWeiScience18}, and 
utilize the equilibrium noise 
expression~\cite{BlanterButtikerPR00,Di_Ventra_book}, $S=4k_BT G$ (with $G$ the conductance),
and the fact that at zero bias the molecular temperature should correspond to that of the contacts,
to introduce a nonequilibrium effective local temperature 
$T_M(A)$ as a function of the noise in Eq.~(\ref{loc_noise_A})
	\begin{equation}
	\label{TM}
	T_M(A)=T_0\,
	\frac{S_{AA}(\omega=0)\lvert_{V_{sd}}}{S_{AA}(\omega=0)\lvert_{0}}.
	\end{equation}
Here, $T_0$ is the temperature in the contacts (assumed to be $300$~K in both $L$ and $R$
reservoirs), $A$ is the surface area of a non-invasive probe that measures the local temperature~\cite{DiVentraRMP11}.

Note that for small surface areas 
($A=6.25\times 10^{-2}$~\AA${}^2$),
over which the integrands in Eq.~(\ref{loc_noise_A})
are constants,  the area size $A$ dependence disappears. This is the case discussed here. 
As a test case we consider the PBDT molecular junction shown in Fig.~\ref{fig1}a.

Figure~\ref{fig2_SM} shows the temperature distribution in the PBDT molecular junction calculated using Eq.~(\ref{TM}) 
at $V_{sd}=1$~V. We assume a non-invasive probe, which measures local noise of the current
projections perpendicular to the probe's surface at a distance of $1.5$~\AA\/ above the molecular
plane. Fig.~\ref{fig2_SM} shows that while the temperature values are of the same order of magnitude,
the temperature {\it profiles} are substantially different for different probe orientations. 
Of course, in realistic measurements the probe is always invasive
and the experimentally measured profiles will mix different contributions. However, such dependence on orientation may be 
observable under certain conditions. 
For example, the effect may be observable in measurements in graphene nano-ribbons. 
This also confirms that the definition of temperature is not unique in a non-equilibrium setting and depends on the 
type of probes used to define it~\cite{DiVentraRMP11, MGANPRB11}.

\section{Conclusion}\label{conclude}
We have introduced the concept, and developed the theory of local-noise spectroscopy as a tool to study transport
properties of systems out of equilibrium. The concept is a natural extension of local fluxes 
which have been used to characterize charge (and energy) flow in nanoscale systems. 

We have shown that the local noise contains rich and complementary (to local fluxes) information on the system. In particular, we have exemplified this tool with the study of bias-induced light emission, intra-system interactions in molecular junctions, and discussed its application to nano-thermometry. 

In the case of light emission we find that outgoing photons
of different frequencies may probe different regions of the molecule, an interesting effect difficult to extract from other probes. 
The cross-correlations of the local noise are instead an indicator of intra-system interactions, and its frequency dependence
yields information on their interaction strength and relevant energy scale. Finally, in the case of nano-thermometry we predict temperature profiles 
dependent on the different probe orientations.

Although LNS may, at the moment, be difficult to realize experimentally, our work shows that it can already be of great help as a theoretical tool to analyze a wide variety of 
physical properties in different non-equilibrium systems. We thus hope that our work will motivate experimentalists in pursuing this line of research that may 
lead to another important spectroscopic probe with the potential to unravel phenomena difficult to detect with other techniques.

\begin{acknowledgments}
M.G. research is supported by the National Science Foundation
(Grant No. CHE-1565939) and the U.S. Department of Energy (Grant No. DE-SC0018201).
\end{acknowledgments}

%

\end{document}